\documentclass[onecolumn,showpacs]{revtex4}

\usepackage{graphicx}
\usepackage{dcolumn}
\usepackage{bm}

\input epsf

\begin{document}

\title{\Large Naked singularity formation for higher dimensional inhomogeneous dust collapse}

\author{\bf Ujjal Debnath}
\email{ujjaldebnath@yahoo.com}
\author{\bf Subenoy Chakraborty}
\email{subenoyc@yahoo.co.in}
 \affiliation{Department of
Mathematics, Jadavpur University, Calcutta-32, India.}

\date{\today}

\begin{abstract}
We investigate the occurrence and nature of a naked singularity
in the gravitational collapse of an inhomogeneous dust cloud
described by higher dimensional Tolman-Bondi space-time for
non-marginally bound case. The naked singularities are found to be
gravitationally strong.
\end{abstract}


\maketitle

\section{\normalsize\bf{Introduction}}
The Cosmic Censorship Conjecture (CCC)[1,2] states that the
space-time singularity produced by gravitational collapse must be
covered by the horizon. The CCC is yet one of the unresolved
problem in General Relativity. In fact the conjecture has not yet
any precise mathematical proof. However the singularity theorems
as such do not state anything about the visibility of the
singularity to an outside observer. Several models related to the
gravitational collapse of matter has so far been constructed
where one encounters a naked singularity [3-8].\\

It is known that the Tolman-Bondi metric admits both naked and
covered singularities depending upon the choice of initial data
and that there is a smooth transition from one phase to the
other. Moreover, according to the strong version of the CCC, such
singularities are not even locally naked, i.e., no non-spacelike
curve can emerge from such singularities (see [2] for reviews of
the CCC). Consequently, examples that appear to violate the CCC
are important and they are an important tools to study this
important issue.\\

It is our purpose here to show that the occurrence of a strong
curvature naked singularity is not confined to self-similar
space-times or null dust only by pointing out a wide class of
Tolman-Bondi models which are non-self-similar in general where
such a singularity forms. In Sec.II, we review 5D Tolman-Bondi
solutions in non-marginally bound case and in Sec.III, we discuss
the nature and existence of central shell focusing naked
singularity. Finally in Sec.IV, we show that gravitational
collapse of 5D space-times gives rise to naked
singularities which are gravitationally strong.\\

\section{\normalsize\bf{Five dimensional Tolman-Bondi solution}}
The five dimensional Tolman-Bondi metric in co-moving co-ordinates
is given by

\begin{equation}
ds^{2}=e^{\nu}dt^{2}-e^{\lambda}dr^{2}-R^{2}d\Omega^{2}_{3}
\end{equation}

where $\nu,\lambda,R$ are functions of the radial co-ordinate $r$
and time $t$ and $d\Omega^{2}_{3}$ represents the metric on the
$3$-sphere. Since we assume the matter in the form of dust, the
motion of particles will be geodesic allowing us to write
~$e^{\nu}=1$. Using comoving co-ordinates one can in view of the
field equations [9], arrive at the following relations in 5
dimensional space-time

\begin{equation}
e^{\lambda}=\frac{R'^{2}}{1+f(r)}
\end{equation}

and

\begin{equation}
\dot{R}^{2}=f(r)+\frac{F(r)}{R^{2}},
\end{equation}

where, $f(r)$ and $F(r)$ are arbitrary functions of radial
co-ordinate $r$ alone with the restriction $1+f(r)>0$ for obvious reasons.
The functions $F(r)$ and $f(r)$ in fact refer to the mass function and the binding
energy function respectively.\\

The energy density $\rho(t,r)$ is therefore given by

\begin{equation}
\rho(t,r)=\frac{3F'(r)}{2R^{3}R'}
\end{equation}

As we are concerned with the gravitational collapse dust we
require $\dot{R}(t,r) < 0$ and without loss of generality we
rescale $R$ such that

\begin{equation}
R(0,r)=r
\end{equation}

Integrating eq.(3) and using the relation (5), we have the
solution

\begin{equation}
R^{2}=r^{2}+f t^{2}-2t\sqrt{F+f r^{2}}
\end{equation}

The central singularity occurs at $r=0$, the corresponding time
being $t=0$. We denote by $\rho(r)$ the initial density:

\begin{equation}
\rho(r)\equiv\rho(0,r)=\frac{3F'}{2r^{3}}~~\Rightarrow
F(r)=\frac{2}{3}\int\rho(r) r^{3} dr
\end{equation}

In 4D case, the mass function $F(r)$ involves the integral
$\int\rho(r) r^{2} dr$ [2]. It can be seen from eq.(4) that the
density diverges faster in 5D as compared to 4D. Thus given a
regular initial surface, the time for the occurrence of the
central shell focusing singularity for the collapse developing
from that surface is reduced as compared to the 4D case. The
reason for this stems from the form of the mass function in
eq.(7). In a ball of radius 0 to $r$, for any given initial
density profile $\rho(r)$, the total mass contained in the ball
is greater than in the corresponding 4D case. Hence there is
relatively more mass-energy collapsing in the space-time as
assumed overall positivity of mass-energy (energy condition).

\section{\normalsize\bf{Existence and nature of naked singularity}}
 In the context of Tolman-Bondi space-times, shell crossing
 singularities are defined by $R'=0, R>0$ and they can be naked. It has
 been shown [10] that shell crossing singularities are
 gravitationally weak and hence such singularities can not be
 considered seriously in the context of the CCC. On the other
 hand, central shell focusing singularities (characterized by $r=0$ and
 $R=0$)are also naked and gravitationally strong as well. Thus,
 unlike shell crossing singularities, shell focusing singularities
 do not admit any metric extension through them. Here we wish to
 investigate a similar situation in our 5D space-time.
 Christodoulou [11] pointed out in the 4D case that the
 non-central singularities are not naked. Hence we shall confine
 our discussion to the central shell focusing singularity.\\

 Now $t=t_{s}(r)$ is the instant of shell focusing singularity
 occurring at $r$ i.e., $R(t_{s}(r),r)=0$. So eq.(6) yields to

\begin{equation}
t_{s}(r)=\frac{\sqrt{f r^{2}+F}-\sqrt{F}}{f}
\end{equation}

Let us assume [12]

\begin{eqnarray}\left.\begin{array}{llll}
F(r)=r^{2}\lambda(r)\\\\
\alpha=\alpha(r)=\frac{r f'}{f}\\\\
\beta=\beta(r)=\frac{r F'}{F}\\\\
R(t,r)=r P(t,r)
\end{array}\right\}
\end{eqnarray}

So using equations (3), (6) and (9), we have the following
expressions

\begin{equation}
R'=\frac{1}{2P}\left[2+\alpha\left(P^{2}-1+\frac{2t}{r}\sqrt{f+\lambda}\right)
-\frac{t}{r}\frac{(\lambda\beta+f\alpha+2f)}{\sqrt{f+\lambda}}\right]
\end{equation}

and

\begin{equation}
\dot{R}'=\frac{1}{2r\sqrt{\lambda+f
P^{2}}}\left[\frac{\lambda}{P^{3}}\left\{2+\alpha\left(P^{2}-1+\frac{2t}{r}\sqrt{f+\lambda}
\right)-\frac{t}{r}\frac{(\lambda\beta+f\alpha+2f)}{\sqrt{f+\lambda}}
\right\}-P f \alpha-\frac{\lambda \beta}{P} \right]
\end{equation}
\\
When $\lambda(r)=$ constant and $f(r)=$ constant, then space-time
becomes self-similar. Now we restrict ourselves to functions
$f(r)$ and $\lambda(r)$ which are analytic at $r=0$, such that
$\lambda(0)>0$ and this implies that $t_{s}(0)=0$. It follows
that the point $r=0, t=0$ corresponds to the central singularity
on the hypersurface $t=0$. From eq.(4) it is seen that the
density at the centre $(r=0)$ behaves with time as
$\rho=\frac{3}{2t^{2}}$. This means that the density is finite at
any time $t$, but becomes singular at $t=0$.\\

We wish to investigate if the singularity, when the central shell
collapses to the centre $r=0$, is naked .The singularity is naked
if and only if there exists a null geodesic that emanates from
the singularity. Let $K^{a}=\frac{dx^{a}}{d\mu}$ be the tangent
vector to the radial null geodesic, where $\mu$ is the affine
parameter. Then we derive the following equations:

\begin{equation}
\frac{dK^{t}}{d\mu}+\frac{\dot{R}'}{\sqrt{1+f}}K^{r}K^{t}=0
\end{equation}

\begin{equation}
\frac{dt}{dr}=\frac{K^{t}}{K^{r}}=\frac{R'}{\sqrt{1+f(r)}}
\end{equation}

Now we introduce a new variable $X=\frac{t}{r}$, then the
function $P(t,r)=P(X,r)$ is given, with the help of (6) and (8),
by

\begin{equation}
[f(X-\Theta)-\sqrt{\lambda}]^{2}=\lambda+f P^{2}
\end{equation}

where we have put $t_{s}(r)=r \Theta(r)$ with

\begin{equation}
P^{2}=1+f X^{2}-2X\sqrt{f+\lambda}
\end{equation}

The nature of the singularity (a naked singularity or a black
hole) can be characterized by the existence of radial null
geodesics emerging from the singularity. The singularity is at
least locally naked if there exist such geodesics and if no such
geodesics exist it is a black hole. If the singularity is naked,
then there exists a real and positive value of $X_{0}$ as a
solution to the equation

\begin{eqnarray}
\begin{array}{c}
X_{0}~=\\
{}
\end{array}
\begin{array}{c}
lim~~~~~ \frac{t}{r}\\
t\rightarrow 0~ r\rightarrow 0
\end{array}
\begin{array}{c}
=~lim~~~~~ \frac{dt}{dr}\\
~~~~~~t\rightarrow 0~ r\rightarrow 0
\end{array}
\begin{array}{c}
=~lim~~~~ \frac{R'}{\sqrt{1+f}}\\
t\rightarrow 0~ r\rightarrow 0
\end{array}
\end{eqnarray}

Define $\lambda_{0}=\lambda(0), \alpha_{0}=\alpha(0), f_{0}=f(0),
\Theta_{0}=\Theta(0)$ and $Q=Q(X)=P(X,0)$, so equations (14) and
(15) reduces to

\begin{equation}
[f_{0}(X-\Theta_{0})-\sqrt{\lambda_{0}}]^{2}=\lambda_{0}+f_{0}
Q^{2}
\end{equation}

and

\begin{equation}
Q^{2}=1+f_{0} X^{2}-2X\sqrt{f_{0}+\lambda_{0}}
\end{equation}

Now from eq.(9) it is to be seen that $\beta(0)=2$. We would
denote $Q_{0}=Q(X_{0})$, the equation (16) simplifies to

\begin{equation}
G(X_{0})=0
\end{equation}

where
\begin{equation}
G(X)=\frac{1}{2Q}\left[2+\alpha_{0}\left(Q^{2}-1+2X\sqrt{f_{0}+\lambda_{0}}\right)
-\frac{(2\lambda_{0}+f_{0}\alpha_{0}+2f_{0})}{\sqrt{f_{0}+\lambda_{0}}}X
-2X Q\sqrt{1+f_{0}} \right]
\end{equation}
\\
If the equation $G(X)=0$ has a real positive root, the singularity
could be naked. If no real positive root is found the
singularity $t=0, r=0$ is obviously black hole.\\

From equation (13), we have

\begin{equation}
\frac{dX}{dr}=\frac{G(X)}{r\sqrt{1+f_{0}}}+\frac{Y(X,r)}{r}
\end{equation}

where $Y(X,r)$ is function of $X$ and $r$ such that at $r=0,
Y(X,0)=0$. Since $X_{0}$ is a root of $G(X)=0$ (see, eq.(19)), so
we can express $G(X)$ in the following form

\begin{equation}
G(X)=q_{0}\sqrt{1+f_{0}}~(X-X_{0})+O(X-X_{0})^{2}
\end{equation}

where
$$
q_{0}=\frac{1}{Q_{0}\sqrt{1+f_{0}}}\left[f_{0}\alpha_{0}X_{0}-\frac{(2\lambda_{0}+
f_{0}\alpha_{0}+2f_{0})}{2\sqrt{f_{0}+\lambda_{0}}}\right]+(\sqrt{f_{0}+\lambda_{0}}-f_{0}X_{0})
\frac{X_{0}}{Q_{0}^{2}}-1
$$

Substituting (22) in (21) we have

\begin{equation}
\frac{dX}{dr}-(X-X_{0})\frac{q_{0}}{r}=\frac{H}{r}
\end{equation}

where $H=H(X,r)=Y(X,r)+O(X-X_{0})$ such that $H(X_{0},0)=0$.\\

Integrating (23) we have

\begin{equation}
X-X_{0}=C r^{q_{0}}+r^{q_{0}}\int H r^{-q_{0}-1} dr
\end{equation}

where $C$ is the constant of integration which labels different geodesics.
From eq.(24) it can be shown that $X\rightarrow X_{0}$ as $r\rightarrow 0$ for
$q_{0}>$ or $<0$. Therefore the single null geodesic described by $C=0$ always
terminates at the singularity $t=0, r=0$ with $X=X_{0}$. For $q_{0}>0$ there are
infinitely many integral curves (characterized by the different values of $C$)
terminate at the singularity. But for $q_{0}<0$ there is only one singular geodesic
(characterized by $C=0$) terminates at the singularity. \\

From (8) we have
\begin{equation}
\Theta_{0}=\frac{1}{f_{0}}(\sqrt{f_{0}+\lambda_{0}}-\sqrt{\lambda_{0}})
\end{equation}

Using (17) to (20) and (25) we have the algebraic equation for
$X_{0}$,

\begin{equation}
a X_{0}^{4}+b X_{0}^{3}+c X_{0}^{2}+d X_{0}+4=0
\end{equation}

where

$$
a=f_{0}\{f_{0}(\alpha_{0}^{2}-4)-4\}~,\hspace{3.4in}
$$
$$
b=8(1+f_{0})\sqrt{f_{0}+\lambda_{0}}-\frac{2\alpha_{0}f_{0}(2f_{0}
+\alpha_{0}f_{0}+2\lambda_{0})}{\sqrt{f_{0}+\lambda_{0}}}
~,\hspace{1.75in}
$$
$$
c=-4(1+f_{0})+4\alpha_{0}f_{0}+\frac{(2f_{0}
+\alpha_{0}f_{0}+2\lambda_{0})^{2}}{f_{0}+\lambda_{0}}~,\hspace{1.95in}
$$
$$
d=-\frac{4(2f_{0}
+\alpha_{0}f_{0}+2\lambda_{0})}{\sqrt{f_{0}+\lambda_{0}}}\hspace{3.3in}
$$
\\

This algebraic equation gives us information about the behavior
of the tangent near the singularity. In fact, the central shell
focusing singularity is at least locally naked if the above
eq.(26) has at least one positive root. The tangents to the
escaping geodesics near the singularity are determined by the
roots of the equation. The smallest root (say, $X_{0}^{s}$) of
$X_{0}$ is called the Cauchy horizon of the space-time as it
indicates the earliest ray escaping from the singularity. Thus no
solution is possible in the region $X_{0}<X_{0}^{s}$. Further,
for non existence of any positive root, it is not possible to
have any future directed radial null geodesic emanating from the
singularity i.e., the singularity is fully covered by trapped
surface and we have only black hole solution.\\

We shall now discuss the nature of the roots of the fourth degree
equation in $X_{0}$. We note that if $a<0$ then equation (26) has
at least one positive and at least one negative root provided
$\alpha_{0}^{2}<4\left(1+\frac{1}{f_{0}}\right)$. For example, if
we choose $f_{0}=1, \lambda_{0}=.1$ and $\alpha_{0}=3$ then
$X_{0}=.354393$ will correspond to Cauchy horizon. If we choose
$a=0$ then the algebraic equation reduces to a cubic equation
which has at least one positive root provided $f_{0}>0$. However,
for $-1<f_{0}<0$ (note that we must have $1+f_{0}>0$, see eq.(2))
at least one positive root is possible if $f_{0}$ is close to
zero, otherwise, for $f_{0}$ close to $-1$ we may or may not have
positive real root. The table shows the numerical results for
different choices of the parameters. Therefore, we conclude that
it is possible to have a naked singularity for various values of
parameters $f_{0}$, $\lambda_{0}$ and $\alpha_{0}$ as shown in the
table.\\\\

TABLE: Roots of equation (26) for various values of the
parameters $f_{0}, \lambda_{0}$ and $\alpha_{0}$.\\

\begin{center}
\begin{tabular}{lcccccc}   \hline
~~~~$f_{0}$ & ~~$\lambda_{0}$    & ~$\alpha_{0}$          &    &    & Roots ($ X_{0}$)\\
\hline\\

$-$.003~&~~~~.034~~~~& ~.2~   & & &~4.32936,~1.16321\\
\\
~$-$.1 &    .11    &  .2    & & & 1.72737,~1.48602\\
\\
~$-$.1 &   .11    &  .5    & & & $-$\\
\\
~$-$.5 &   .51    &  .001   & & & $-$\\
\\
~~~~1 &   .1    &  3    & & & 10.3982,~1.48206,~.354393\\
\\
~~~~3 &   .001    &  .1    & & & .557735,~.481183\\
\\
~~~~9 &   .001    &  1    & & & .329341\\
\\
~~~~9 &   .01    &  5    & & & .093828\\
\\

\hline\\

\end{tabular}
\end{center}

\section{\normalsize\bf{Strength of naked singularity}}
 Finally we need to determine curvature strength of the naked
 singularity which is an important aspect of a singularity [13].
 There has been attempt to relate the strength of a singularity to
 its stability [14]. A singularity is gravitationally strong or
 simply strong if it destroys by crushing or stretching any
 objects that fall into it and weak if no object that falls into
 the singularity is destroyed in this way. Clarke and Krolak [15]
 have shown that a sufficient condition for strong curvature
 singularity defined by Tipler [16] is that for at least one
 non-spacelike geodesic with affine parameter $\mu$, in the limiting
 approach to the singularity, we must have

\begin{equation}
\begin{array}{c}
lim\\
\mu\rightarrow 0\\
\end{array}
\begin{array}{c}
\mu^{2}\psi=\\
{}
\end{array}
\begin{array}{c}
lim\\
\mu\rightarrow 0\\
\end{array}
\begin{array}{c}
\mu^{2} R_{a b}K^{a}K^{b}>0\\
{}
\end{array}
\end{equation}

where $R_{a b}$ is the Ricci tensor. Our purpose here is to
investigate the above condition along future directed radial null
geodesics that emanates from the naked singularity. Now equation
(27) can be expressed as

\begin{equation}
\begin{array}{c}
lim\\
\mu\rightarrow 0\\
\end{array}
\begin{array}{c}
\mu^{2}\psi=\\
{}
\end{array}
\begin{array}{c}
lim\\
\mu\rightarrow 0\\
\end{array}
\begin{array}{c}
\frac{3F'}{2r P R'}\left(\frac{\mu K^{t}}{R}\right)^{2}\\
{}
\end{array}
\end{equation}

Using L'Hospital's rule and using equations (10) and (11), the
equation (28) can be written as

\begin{equation}
\begin{array}{c}
lim\\
\mu\rightarrow 0\\
\end{array}
\begin{array}{c}
\mu^{2}\psi=\\
{}
\end{array}
\frac{12\lambda_{0}Q_{0}X_{0}(\lambda_{0}+f_{0}Q_{0}^{2})\sqrt{1+f_{0}}}
{\left[2\lambda_{0}X_{0}\sqrt{1+f_{0}}-2\lambda_{0}Q_{0}-f_{0}\alpha_{0}Q_{0}^{3}\right]^{2}} >0\\
\end{equation}

Thus along radial null geodesics coming out of a singularity
$\begin{array}{c}
lim\\
\mu\rightarrow 0\\
\end{array}
\begin{array}{c}
\mu^{2}\psi\\
{}
\end{array}$ is finite and hence the strong curvature condition is
satisfied.\\

\section{\normalsize\bf{Concluding Remarks}}
 In this paper, the 5D Tolman-Bondi space-time have been studied
 for the formation of naked singularities in spherical dust
 collapse in non-marginally bound case ($f\neq 0$). The nature of
 the singularity depends on the value of the curvature
 parameter $f$ at $r=0$ (i.e. $f_{0}$). In fact for $f_{0}>0$ or
 $f_{0}<0$ (but close to zero) the singularity may be naked but it
 depends also  on the choice of the other two parameters involved. The
 strong curvature condition for naked singularity is satisfied for
 this 5D inhomogeneous dust collapse. The natural questions that
 arise how the extra dimension influence the formation of a naked
 singularity and what is the role of the curvature parameter in
 classifying the singularity. These questions will be addressed in
 a subsequent paper. For future work, it will be interesting to
 study the exact restrictions for formation of naked singularity
 on the parameters involved and their physical implications.\\

{\bf Acknowledgement:}\\

The authors are thankful to the members of Relativity and
Cosmology Research Centre, Department of Physics, Jadavpur
University for helpful discussion. One of the  authors (U.D) is
thankful to CSIR (Govt. of India) for awarding a Junior Research Fellowship.\\

{\bf References:}\\
\\
$[1]$  R. Penrose, {\it Riv. Nuovo Cimento} {\bf 1} 252 (1969);
in General Relativity, an Einstein Centenary Volume, edited by S.W.
Hawking and W. Israel (Cambridge Univ. Press, Cambridge, 1979).\\
$[2]$  P.S. Joshi, {\it Global Aspects in Gravitation and Cosmology}
(Oxford Univ. Press, Oxford, 1993).\\
$[3]$  B. Waugh and K. Lake, {\it Phys. Rev. D} {\bf 38} 1315 (1988).\\
$[4]$  J. P. S. Lemos, {\it Phys. Lett. A} {\bf 158} 271 (1991);
{\it Phys. Rev. Lett.} {\bf 68} 1447 (1992).\\
$[5]$  P. S. Joshi and I. H. Dwivedi, {\it Class. Quantum Grav.}
{\bf 16} 41 (1999).\\
$[6]$  A. IIha and J. P. S. Lemos, {\it Phys. Rev. D} {\bf 55}
1788 (1997).\\
$[7]$  A. IIha, A. Kleber and J. P. S. Lemos, {\it J. Math.
Phys.} {\bf 40} 3509 (1999).\\
$[8]$  J. F. V. Rocha and A. Wang, {\it ibid} {\bf 17} 2589
(2000).\\
$[9]$  S. G. Ghosh and A. Beesham, {\it Class. Quantum Grav.}
{\bf 17} 4959 (2000).\\
$[10]$  R. P. A. C. Newman, {\it Class. Quantum Grav.}
{\bf 3} 527 (1986).\\
$[11]$  D. Christodoulou, {\it Commum. Math. Phys.}
{\bf 93} 171 (1984).\\
$[12]$  I. H. Dwivedi and P. S. Joshi, {\it Class. Quantum Grav.}
{\bf 9} L69 (1992).\\
$[13]$  F. J. Tipler, {\it Phys. Lett. A} {\bf 64}
8 (1987).\\
$[14]$  S. S. Deshingkar, P. S. Joshi and I. H. Dwivedi, {\it
Phys. Rev. D} {\bf 59} 044018 (1999).\\
$[15]$  C. J. S. Clarke and A. Krolak, {\it J. Geom. Phys.} {\bf
2} 127 (1986).\\
$[16]$  F. J. Tipler, C. J. S. Clarke and G. F. R. Ellis, General
Relativity and Gravitation ed. A Held (New York, Plenum) 1980.\\

\end{document}